\documentclass[10pt,twocolumn,romanappendices]{IEEEtran}
\usepackage{cite}
\usepackage{graphicx, amsfonts,amsmath, enumerate, amsthm, color,amssymb,etaremune,array,url}
\usepackage{setspace,multirow,psfrag}
\usepackage{graphicx}
\usepackage{subcaption}
\graphicspath{{../figures/}}

\newtheorem{scenario}{Scenario}
\newtheorem{proposition}{Proposition}

\begin{document}
\title{Does Massive MIMO Fail in Ricean Channels?}
\author{Michail Matthaiou,~\IEEEmembership{Senior Member,~IEEE,} Peter J. Smith,~\IEEEmembership{Fellow,~IEEE,} \\Hien Quoc Ngo,~\IEEEmembership{Member,~IEEE,} and Harsh Tataria,~\IEEEmembership{Member,~IEEE}

\thanks{

Manuscript received February 16, 2018; revised May 10, 2018; accepted July 1, 2018. Date of publication July xxx, 2018; date of current version
xxx xxx, 2018. The associate editor coordinating the review of this paper and approving it for publication was J. Lee.

M. Matthaiou, H. Q. Ngo and  H. Tataria are with the Institute of Electronics, Communications and Information Technology (ECIT), Queen's University, Belfast, Belfast BT3 9DT, U.K. (e-mail: m.matthaiou@qub.ac.uk).

P. J. Smith is with the School of Mathematics and Statistics, Victoria University of Wellington, Wellington 6140, NZ (e-mail: peter.smith@vuw.ac.nz).

The work of M. Matthaiou and H. Tataria was supported by EPSRC, UK, under grant EP/P000673/1. The work of P. J. Smith was supported
by the Royal Academy of Engineering, UK, via the Distinguished Visiting Fellowship DVF1617/6/29.}}
\maketitle

\begin{abstract}
Massive multiple-input multiple-output (MIMO) is now making its way to the standardization exercise of future 5G networks. 
Yet, there are still fundamental questions pertaining to the robustness of massive MIMO against physically detrimental propagation conditions. On these grounds, we identify scenarios under which massive MIMO can potentially fail in Ricean channels, and characterize them physically, as well as, mathematically. Our analysis extends and generalizes a stream of recent papers on this topic and articulates emphatically that such harmful scenarios in Ricean fading conditions are unlikely and can be compensated using any standard scheduling scheme. This implies that massive MIMO is intrinsically effective at combating 
inter-user interference and, if needed, can avail of the base-station scheduler for further robustness. 
\end{abstract}


%Digital Object Identifier XXXXX/TSP.2010.09945
\IEEEpeerreviewmaketitle

\section{Introduction}
Massive multiple-input multiple-output (MIMO) is nowadays a well-established technology which forms the backbone of the fifth-generation (5G) \cite{MAMI_book}. The seamless development of massive MIMO since 2010 has been based on the concept of favorable propagation \cite{Ngo_EUSIPCO}, which leverages asymptotic properties of Gaussian random vectors. That is, as the number of base-station (BS) antennas becomes unconventionally large, channel vectors become pairwise orthogonal. In order for this key property to hold, a common assumption is that the individual channel vectors have independent, zero-mean Gaussian entries with a particular finite variance. 

Surprisingly, once we start moving away from these conditions little is known about the massive MIMO performance.
%The early work of \cite{Zhang} derived exact and asymptotic results for massive MIMO in Ricean fading conditions. 
In \cite{Ngo_EUSIPCO}, it was shown that, for a fixed number of users and under pure line-of-sight (LoS) conditions, the orthogonality between two random channel vectors breaks whenever their angular difference scales as ${\cal O}(1/M)$, where $M$ is the number of BS antennas. The authors of \cite{Masouros} investigated the performance of massive MIMO when the total electrical length of the BS antenna array is fixed. Their results showcased that inter-user interference for pure LoS conditions does not vanish in the massive MIMO regime. In \cite{Emil_PC}, Bj\"{o}rnson \textit{et al.} proved that for correlated Rayleigh fading conditions, if the pilot-sharing users have asymptotically linearly independent covariance matrices, the massive MIMO capacity grows without bound. \color{black}
The work of \cite{Luca} leveraged tools of random matrix theory to derive asymptotic expressions for the average rate and concluded that Ricean fading is more beneficial than Rayleigh fading. \color{black} Finally, \cite{Beualieu} derived conditions to guarantee favorable propagation for different array topologies. 

We herein consider a far more general propagation scenario compared to \cite{Ngo_EUSIPCO, Masouros, Emil_PC, Luca, Beualieu}, which is modeled via the semi-correlated Ricean distribution, where each user has a different covariance matrix and a different $K$-factor. This general model is inherently suitable for future dense networks, where different sets of incident directions are likely to be observed by geographically separated terminals. These propagation conditions cause variations in the covariance profiles across different users \cite{Caire}. Then, we identify, both physically and mathematically, scenarios under which the mean inter-user interference power with maximum-ratio processing does not vanish in the large-antenna limit, thereby limiting the seemingly extensive massive MIMO gains. Our performance metric is the mean interference power since it provides an average over the small-scale fading making possible  a second-order characterization of a multi-user massive MIMO system. Our analysis provides a number of important observations, for this class of Ricean fading channels: (a) It is unlikely that massive MIMO will fail; (b) Failure occurs when we have strong alignment of two distinct LoS responses and/or non-vanishing alignment of a LoS response of the $k$-th user with the eigenvectors of the covariance matrix of the $\ell$-th user; (c) If any of these scenarios kicks in, a standard scheduling scheme can remove the undesired user(s) from the communication link, hence, minimizing the inter-user interference; (d) under some mild conditions, the instantaneous Gram matrix normalized by $M$ converges to its mean in the mean-square sense, respectively.

\textit{Notation:} We use upper and lower case boldface to denote matrices and vectors, respectively. The $n\times n$ identity matrix is expressed as $\mathbf{I}_{n}$. A complex normal vector with mean ${\mathbf b}$ and covariance ${ \bf \Sigma}$ reads as $\mathcal{CN}({ \mathbf b},{ \bf \Sigma})$. The expectation of a random variable is denoted as ${\mathbb E}\left[\cdot \right] $, while the matrix trace by $\mathrm{{tr}(\cdot )}$. The symbols $\left( \cdot \right) ^{T}$ and $\left(\cdot \right) ^{H}$ represent the transpose and Hermitian transpose of a matrix.

\vspace{-14pt}
\section{System Model}
Consider an uplink cellular system with $M$ BS antennas which serve $L$ single-antenna users in the same time-frequency resource with $M\gg L$. 
%The $M\times 1$ received signal vector, ${\bf y}$ at the BS is given by
%\begin{align}
	%{\bf y} = \sqrt{p_u}{\bf GD}^{1/2}{\bf s}+ {\bf n} 
	%\label{eq:rec_vector}
%\end{align}
%where $p_u$ is the transmit power, ${\bf s} = [s_1, s_2,\ldots,s_L]^T$ is the transmitted signal vector whose entries satisfy ${\mathbb E}[|s_i|^2]=1, i=1,\ldots,L$, while ${\bf n}=[n_1,n_2,\ldots,n_M]^T$ is the additive white Gaussian noise vector with $n_m\sim{\cal CN}(0,\sigma_m^2), m=1,\ldots,M$. The $L\times L$ diagonal matrix ${\bf D}$ accounts for large-scale fading effects which will be discussed later in the paper. 
We focus on a very general fading scenario with semi-correlated Ricean fading, where each user has a different $K$-factor and a semi-positive definite $M\times M$ covariance matrix ${\bf R}_k$, with $\mathrm{tr}({\bf R}_k)=M$. \color{black}In the subsequent theoretical analysis, we temporarily ignore the effects of large-scale fading as they do not affect our findings and, most importantly, because we are exclusively interested in the macroscopic impact of small-scale fading. Yet, in the simulation results, we do consider a realistic large-scale fading model. \color{black} The $M\times 1$ channel from the $k$-th user to the BS is
\begin{align}
	{\bf g}_k & = {\sqrt{\frac{K_k}{K_k+1}}\bar{\bf {h}}_k}+{\sqrt{\frac{1}{K_k+1}}{\bf R}_k^{1/2}\widetilde{\bf {h}}_k}, \label{eq:System_model1} 
\end{align}
where $K_k$ is the Ricean $K$ factor, $\bar{\bf {h}}_k$ is the deterministic LoS component with $||\bar{\bf {h}}_k||^2=M$ and 
$\widetilde{\bf {h}}_k\sim {\cal CN} ({\bf 0},{\bf I}_M)$ models the diffuse multipath components. 

With maximum-ratio processing and \color{black} perfect channel state information (CSI)\footnote{\color{black} Interestingly, the case of perfect CSI is mathematically analogous to the case of imperfect CSI with orthogonal pilot signaling with the only difference pertaining to the covariance matrix of the estimate, which is a shifted version of ${\bf R}_k$. For this reason and to keep the notation neat, we work with the former case.}, \color{black} the interference created by the $l$-th to the $k$-th user is defined as:
$	T_{k\ell}\triangleq|{\bf g}_k^H{\bf g}_\ell|^2$, such that the total interference power seen by the $k$-th user is equal to
$\sum\limits_{\ell=1,\ell\neq k}^L |{\bf g}_k^H{\bf g}_\ell|^2$. Our performance metric will be the mean interference power \cite[Lemma 2]{Tataria2017}:
\begin{align}
	&{\mathbb E}\left[T_{k\ell}\right] = \underbrace{{\frac{K_\ell}{K_\ell+1}}\frac{1}{K_k+1}\big(\bar{\bf {h}}_\ell^H{\bf R}_k	\bar{\bf {h}}_\ell \big)}_{{\tt term1}}+\underbrace{\frac{{\rm tr}({\bf R}_\ell{\bf R}_k)}{(K_k+1)(K_\ell+1)}}_{{\tt term2}}\nonumber\\
	&\hspace{-0.1in}+\hspace{-0.04in}\underbrace{\frac{K_k}{K_k+1}\frac{K_\ell}{K_\ell+1}|\bar{\bf {h}}_\ell^H\bar{\bf {h}}_k|^2}_{{\tt term3}}\hspace{-0.03in}+\hspace{-0.03in}\underbrace{{\frac{K_k}{K_k+1}}\frac{1}{K_\ell+1}\big(\bar{\bf {h}}_k^H{\bf R}_\ell	\bar{\bf {h}}_k \big)}_{{\tt term4}}.
	\label{eq:mean_power}
\end{align}
In the following sections, we will separately study these four individual terms in \eqref{eq:mean_power} and, in particular, their scaling behavior with an increasing number of antennas $M$.
%\footnote{It is interesting to note that analyzing the instantaneous interference term $T_{k\ell}$ in the limit of $M \rightarrow \infty$ will lead to precisely the same mathematical conclusions at the expense of much heavier notation.} 
Our main objective is to identify physical scenarios under which any of these four terms scales as ${\cal O}(M^2)$, which is also the scaling order of the desired signal power \cite{MAMI_book, Masouros}.

\vspace{-10pt}
\section{When does massive MIMO fail?}
\subsection{Analysis of ${{\tt term1}}$ in \eqref{eq:mean_power}}
To analyze this term, we focus our attention on the quadratic form inside ${{\tt term1}}$, that is, $\bar{\bf {h}}_\ell^H{\bf R}_k	\bar{\bf {h}}_\ell$. This term is a quadratic form of the LoS vectors $\bar{\bf {h}}_\ell$ and the covariance matrix of user $k$, ${\bf R}_k$.\footnote{Note that a similar analysis can be pursued for ${{\tt term4}}$ in \eqref{eq:mean_power}, which is omitted for the sake of brevity.} We can now sort the $M$ real eigenvalues of ${\bf R}_k$ in descending order as follows $\lambda_1^{(k)}\geq \lambda_2^{(k)}\geq\ldots\geq\lambda_M^{(k)}\geq 0$ with $\sum_{i=1}^M \lambda_i^{(k)} = {\rm tr}\big({\bf R}_k\big) = M, \forall k=1,\ldots,L$. Hence, the eigenvalue decomposition of ${\bf R}_k$ reads as ${\bf R}_k = {\bf U}_k{\bf \Lambda}_k{\bf U}_k^H$, where
${\bf U}_k\triangleq\left[{\bf u}_1^{(k)},{\bf u}_2^{(k)},\ldots,{\bf u}_M^{(k)}\right]$ is a unitary matrix that contains the eigenvectors of ${\bf R}_k $ and ${\bf \Lambda}_k \triangleq {\rm diag}\left(\lambda_1^{(k)},\lambda_2^{(k)},\ldots, \lambda_M^{(k)}\right)$. 

We now have that
\begin{align}
	\frac{1}{M^2}\bar{\bf {h}}_\ell^H{\bf R}_k	\bar{\bf {h}}_\ell = \frac{1}{M}\frac{\bar{\bf {h}}_\ell^H}{\sqrt{M}}{\bf R}_k\frac{\color{black}{\bar{\bf {h}}_\ell} \color{black}}{\sqrt{M}} = 
	 \frac{1}{M}\frac{\bar{\bf {h}}_\ell^H}{|| \bar{\bf {h}}_\ell ||}{\bf R}_k\frac{\color{black}\bar{\bf {h}}_\ell \color{black}}{|| \bar{\bf {h}}_\ell ||}
	%\leq \frac{1}{M} \lambda_{\rm{max}}({\bf R}_k)
\end{align}
which from the Rayleigh-Ritz theorem can be lower and upper bounded as follows
\begin{align}
	\frac{\lambda_M^{(k)}}{M}\leq\frac{1}{M^2}\bar{\bf {h}}_\ell^H{\bf R}_k	\bar{\bf {h}}_\ell \leq \frac{\lambda_1^{(k)}}{M}\leq 1.
	\label{eq:Ritz}
\end{align}
From \eqref{eq:Ritz}, it is clear that situations exist where $\bar{\bf {h}}_\ell^H{\bf R}_k	\bar{\bf {h}}_\ell$ scales
as ${\cal O}(M^2)$. One such case is the extreme (and unlikely) situation where $\bar{\bf {h}}_\ell$ is aligned
with the weakest eigenvector of ${\bf R}_k$ and $\lambda_M^{(k)}$ is ${\cal O}(M)$.
A set of milder conditions under which channel orthogonality breaks is considered below.
%We point out that the lower bound in \eqref{eq:Ritz}, $\lambda_M^{(k)}/M$, is achieved whenever $\bar{\bf {h}}_\ell$ is aligned with the weakest eigenvector of ${\bf R}_k$. In such an extreme (and unlikely) case, ${{\tt term1}}/M^2$ will, of course, not vanish in the massive MIMO regime if $\lambda_M^{(k)}$ is ${\cal O}(M)$. We will now  further investigate this interesting scenario and provide a set of milder conditions under which the channel orthogonality breaks down. 
\begin{scenario}
Note that by definition $\bar{\bf {h}}_\ell$ can be expressed as a linear combination 
of the linearly independent eigenvectors ${\bf u}_i^{(k)}$. If the corresponding eigenvalue(s) scale as
${\cal O}(M)$, then ${{\tt term1}}/M^2$ does not vanish as $M\rightarrow\infty$.
\begin{proof}
We express $\bar{\bf {h}}_\ell$ as a linear combination of the eigenvectors ${\bf u}_i^{(k)}, i=1,\ldots,M$ such that $\frac{1}{\sqrt{M}} \bar{\bf {h}}_\ell = \sum_{i=1}^M \beta_i {\bf u}_i^{(k)}$, where $\sum_{i=1}^M |\beta_i|^2 = 1$. Thus, we have 
\begin{align}
&\frac{1}{M^2}\bar{\bf {h}}_\ell^H{\bf R}_k	\bar{\bf {h}}_\ell = \frac{1}{M} \sum_{i=1}^M \beta_i^*\left({\bf u}_i^{(k)}\right)^H\left({\bf U}_k{\bf \Lambda}_k{\bf U}_k^H\right)\sum_{j=1}^M  \beta_j{\bf u}_j^{(k)} \nonumber \\
&= \frac{1}{M} \sum_{i=1}^M \beta_i^*\left(0,\ldots,1,\ldots,0\right){\bf \Lambda}_k{\bf U}_k^H\sum_{j=1}^M  \beta_j{\bf u}_j^{(k)} \nonumber\\
& = \frac{1}{M} \sum_{i=1}^M \beta_i^*\lambda_i^{(k)}\beta_i = \frac{1}{M} \sum_{i=1}^M |\beta_i|^2\lambda_i^{(k)}. 
\label{eq:Lin_comb}
\end{align}
Thus, \eqref{eq:Lin_comb} does not converge to zero if ${\bf {h}}_\ell$ has non-vanishing alignment $\left(|\beta_i|^2>0~{\rm as}~M\rightarrow\infty\right)$ with one or more eigenvectors, ${\bf u}_i^{(k)}$, whose eigenvalues, $\lambda_i^{(k)}$, scale as ${\cal O}(M)$. 
\end{proof}
\end{scenario}
\textit{Discussion:} Fundamentally, ${{\tt term1}}/M^2$ will not vanish if three conditions are fulfilled: (a) ${\bf R}_k$ has one or more eigenvalues that scale as ${\cal O}(M)$; (b) $\bar{\bf {h}}_\ell$ must align with the corresponding eigenvectors or any linear combination of them; (c) this alignment should be preserved as $M \rightarrow \infty$. Note that for full-rank ${\bf R}_k$, all $\lambda_i^{(k)}$ are de facto positive and this increases the chances of non-vanishing alignment between the LoS response $\bar{\bf {h}}_\ell$ and 
the eigenvectors ${\bf u}_i^{(k)}$. On the contrary, if ${\bf R}_k$ is rank-deficient with rank $r$, then ${{\tt term1}}/M^2$ will vanish unless we have non-vanishing alignment of $\bar{\bf {h}}_\ell$ with ${\bf u}_i^{(k)}, i=1,..,r$,  which is again an improbable situation to kick in. Interestingly, \cite{Emil_PC} articulated that rank-deficient covariance matrices with orthogonal support eliminate pilot contamination resulting in unbounded massive MIMO capacity.

%We point out that conditions (b) and (c) are easier to satisfy since for full-rank ${\bf R}_k$, we have that 
%${\bf u}_1^{(k)},\ldots,{\bf u}_M^{(k)}$ span $\mathbb{C}^M$ and, thus, some $\beta_i$'s must be \textsl{de facto} non-zero. 
%On the contrary, if ${\bf R}_k$ is rank-deficient, then it will be possible for $\bar{{\bf {h}}}_\ell \notin {\rm span}\left({\bf U}_k\right)$, thereby making ${{\tt term1}}/M^2$ vanish for large $M$. 

\vspace{-10pt}
\subsection{Analysis of ${{\tt term2}}$ in \eqref{eq:mean_power}}
To analyze the behavior of this term, we first recall the decomposition of ${\bf R}_{k} = {\bf U}_k{\bf \Lambda}_k{\bf U}_k^H$.
Then, we have for the trace term inside ${{\tt term2}}$:
\begin{align*}
	&\frac{{\rm tr}({\bf R}_\ell{\bf R}_k)}{M^2}= \frac{{\rm tr}\left({\bf \Lambda}_k^{1/2}{\bf U}_k^H{\bf R}_\ell{\bf U}_k{\bf \Lambda}_k^{1/2}\right)}{M^2}\nonumber\\
	&=\frac{1}{M^2}\sum_{i=1}^M  \lambda_i^{(k)}{\left({\bf u}_i^{(k)}\right)^H{\bf R}_\ell{\bf u}_i^{(k)}}\leq \lambda_1^{(\ell)} \sum_{i=1}^M  \frac{\lambda_i^{(k)}}{M^2}  = \frac{\lambda_1^{(\ell)}}{M}\leq 1 
	%\label{eq:corr_trace}
	\end{align*}
where $\lambda_1^{(\ell)}$ is the maximum eigenvalue of ${\bf R}_\ell$. Note that the upper bound above, ${\lambda_1^{(\ell)}}/{M}$, is achieved when all eigenvectors of ${\bf R}_k$ align with the principal eigenvector of ${\bf R}_\ell$.
Although this scenario is mathematically possible, it is highly unrealistic in practice. We will now delineate the general conditions under which ${{\tt term2}}/M^2$ does not vanish as $M\rightarrow\infty$.
\begin{scenario}
By definition ${\bf u}_i^{(\ell)}$ can be expressed as a linear combination 
of the linearly independent eigenvectors ${\bf u}_i^{(k)}$ (or vice versa). If the corresponding eigenvalue(s) scale as ${\cal O}(M)$, then ${{\tt term2}}/M^2$ does not vanish as $M\rightarrow\infty$.
\begin{proof}
We simply replace $\bar{\bf {h}}_\ell$ with ${\bf u}_i^{(\ell)}$ in Scenario 1.
\end{proof}
\end{scenario}
%\textit{Discussion:} 
This scenario requires non-vanishing alignment of the eigenvectors ${\bf u}_i^{(\ell)}$ with the eigenvectors ${\bf u}_i^{(k)}$ and the corresponding eigenvalue(s) of any of ${\bf R}_\ell, {\bf R}_k$ to be scaling as ${\cal O}(M)$. %This scenario is again highly unlikely to cause a problem. 

%Now, let us consider an extreme scenario of fully-aligned covariance matrices, i.e., 
%${\bf {R}}_\ell = \eta {\bf {R}}_k$, where $\eta\in\mathbb {C}$, we have that
%\begin{align}
	%\frac{{\rm tr}({\bf R}_\ell{\bf R}_k)}{M^2} = \frac{\eta{\rm tr}\left({\bf U}_k{\bf \Lambda}_k^2{\bf U}_k^H\right)}{M^2}=
	%\frac{\eta}{M^2}\sum_{i=1}^M  \Big(\lambda_i^{(k)}\Big)^2\leq \eta
%\end{align}
%with equality if only and if $\lambda_1^{(k)}=M$ and $\lambda_2^{(k)}=\ldots=\lambda_M^{(k)}=0$ (i.e. ${\bf R}_k$ is rank-1 matrix). 
%Note that this extreme case only needs $\lambda_1^{(k)}$ to scale ${\cal O}(M)$ to cause a problem.  

\vspace{-10pt}
\subsection{Analysis of ${{\tt term3}}$ in \eqref{eq:mean_power}}
In \eqref{eq:mean_power}$, {{\tt term3}}$ represents the amount of cross-interference between two LoS vectors. We now identify two scenarios that make this term have asymptotically non-vanishing power.
\begin{scenario}
When the two LoS vectors $\bar{\bf {h}}_k$ and $\bar{\bf {h}}_\ell$ are aligned, ${{\tt term3}}/M^2$ does not vanish 
as $M\rightarrow\infty$.
\begin{proof}
Assuming that $\bar{\bf {h}}_k=\alpha\bar{\bf {h}}_\ell$, 
where $|\alpha|^2=1$, we have  
\begin{align*}
	&\frac{1}{M^2}\frac{K_k}{K_k+1}\frac{K_\ell}{K_\ell+1}|\bar{\bf {h}}_\ell^H\bar{\bf {h}}_k|^2 = 
	 \frac{K_k}{K_k+1}\frac{K_\ell}{K_\ell+1} \qedhere.
\end{align*}
\end{proof}
\end{scenario}

\begin{scenario}
Let us now consider the practical scenario where the BS is equipped with a uniform linear array (ULA). This setup was also investigated in \cite{Ngo_EUSIPCO, GLOBECOM}. When the angular difference between $\bar{\bf {h}}_k$ and $\bar{\bf {h}}_\ell$ scales as ${\cal O}(1/M^c)$, with $c\geq1$ ${{\tt term3}}/M^2$ does not vanish in the massive MIMO regime. 
\begin{proof}
In this case, the LoS vector $\bar{\bf {h}}_k$ can be expressed:
\begin{align}
	\bar{\bf {h}}_k = \left[1,e^{-j\frac{2\pi d}{\lambda}\sin{(\theta_k)}}, \cdots,e^{-j\frac{2\pi d}{\lambda} (M-1)\sin{(\theta_k)}}\right]^T.\label{eq:st_vec}
\end{align}

We now assume that $\sin{(\theta_\ell)}-\sin{(\theta_k)}=\gamma/M$ where $\gamma \in \mathbb{R}^+$. \color{black}Then, we can show using the technique of \cite{Ngo_EUSIPCO} that
\begin{align*}
	&\hspace{-20pt}\frac{1}{M^2}\frac{K_k}{K_k+1}\frac{K_\ell}{K_\ell+1}|\bar{\bf {h}}_\ell^H\bar{\bf {h}}_k|^2\nonumber\\
	%&\hspace{-20pt}= \frac{1}{M^2}\frac{K_k}{K_k+1}\frac{K_\ell}{K_\ell+1}\left|\sum_{m=0}^{M-1} e^{jm\frac{2\pi d}{\lambda}(\sin{(\theta_\ell)-\sin{(\theta_k))}}}\right|^2 \nonumber \\
	%&\hspace{-20pt}=\frac{1}{M^2}\frac{K_k}{K_k+1}\frac{K_\ell}{K_\ell+1}
	%\left|\frac{1-e^{jM\frac{2\pi d}{\lambda}(\sin{(\theta_\ell)-\sin{(\theta_k))}}}}{1-e^{j\frac{2\pi d}{\lambda}(\sin{(\theta_\ell)-\sin{(\theta_k))}}}}\right|^2\\
	&\hspace{-20pt}\rightarrow \frac{K_k}{K_k+1}\frac{K_\ell}{K_\ell+1} \left(\frac{\lambda}{2\pi \gamma d}\right)^2\left| e^{j\frac{2\pi \gamma d}{\lambda}-1}\right|^2,~\textrm{as}~M\rightarrow\infty \qedhere.
\end{align*}
\end{proof}
\end{scenario}

\color{black}
\textit{Discussion:} These two scenarios showcase that whenever the LoS vectors are either (a) aligned in the complex plane or (b) have similar angular characteristics, the channel orthogonality breaks down. As a matter of fact, the higher the values of $\alpha$ and $\gamma$ are, the further away from favorable propagation we move. Interestingly, Scenario 4 is a special case of Scenario 3, since it requires only correlated angular characteristics. Note that stronger LoS conditions (i.e. higher $K$-factors) will only make things even worse as it will be extremely difficult to discriminate any two channel vectors $\bar{\bf {h}}_k$ and $\bar{\bf {h}}_\ell$. 

\vspace{-10pt}
\subsection{Implications}
Putting everything together, we conclude that massive MIMO will not fail if these mild conditions are fulfilled:
\renewcommand{\labelenumi}{C\arabic{enumi})}
%\begin{etaremune}
%\item 
\begin{align}
	&{\rm C1}: \frac{\bar{\bf {h}}_\ell^H{\bf R}_k	\bar{\bf {h}}_\ell}{M^2}\rightarrow 0,~~\textrm{as}~~M \rightarrow  \infty,~\forall k,\ell = 1,\ldots,L\\
	&{\rm C2}: \frac{{\rm tr}({\bf R}_\ell{\bf R}_k)}{M^2}\rightarrow 0,~\textrm{as}~~M \rightarrow  \infty,~\forall k,\ell = 1,\ldots,L\\
	&{\rm C3}:  \frac{|\bar{\bf {h}}_\ell^H\bar{\bf {h}}_k|^2}{M^2}\rightarrow 0,~~~\textrm{as}~~M \rightarrow  \infty,~\forall k,\ell = 1,\ldots,L.
\end{align}
We can now leverage the conditions above to present the following result that is very useful for the performance analysis of massive MIMO. For this analysis, we need to define
${\bf G}\triangleq[{\bf g}_1,{\bf g}_2,\ldots,{\bf g}_L]\in {\mathbb C}^{M\times L}$.
\begin{proposition}
If the conditions C1 and C2 are fulfilled, then the instantaneous Gram matrix ${\bf G}^H{{\bf G}}$ converges as follows
\begin{align}
\frac{1}{M}{\bf G}^H{{\bf G}}\stackrel{{\rm m.s.}}{\longrightarrow} \frac{1}{M}{\mathbb E}\left[{\bf G}^H{\bf G} \right] 
\end{align}
where $\stackrel{{\rm m.s.}}{\longrightarrow}$ denotes convergence in the mean-square sense.
\begin{proof}
The $(k,\ell)$-th element of ${\bf G}^H{{\bf G}}$ can be expressed as
\begin{align*}
	&\frac{1}{M}\left[{\bf G}^H{{\bf G}}\right]_{k\ell} =\frac{1}{M}\bigg({\sqrt{\frac{K_k}{K_k+1}}\sqrt{\frac{K_\ell}{K_\ell+1}}\bar{\bf {h}}_k^H\bar{\bf {h}}_\ell} \nonumber\\
	&+{\sqrt{\frac{K_k}{K_k+1}}\sqrt{\frac{1}{K_\ell+1}}\bar{\bf {h}}_k^H{\bf R}_\ell^{1/2}\widetilde{\bf {h}}_\ell}\hspace{40pt}{\rm (S_1)} \nonumber\\
	&+ \sqrt{\frac{1}{K_k+1}}\sqrt{\frac{K_\ell}{K_\ell+1}}\widetilde{\bf {h}}_k^H{\bf R}_k^{1/2}\bar{\bf {h}}_\ell\hspace{40pt}{\rm (S_2)} \nonumber\\
	&+{\sqrt{\frac{1}{K_k+1}}\sqrt{\frac{1}{K_\ell+1}}\widetilde{\bf {h}}_k^H{\bf R}_k^{1/2}{\bf R}_\ell^{1/2}\widetilde{\bf {h}}_\ell}\bigg)\hspace{12pt}{\rm (S_3)}.  
\end{align*}
It is easy to see that for the term ${\rm S_1}$, we have that
\begin{align}
	{\mathbb E}\left[{\rm S_1}\right] &= 0\\
	{\mathbb E}\left[|{\rm S_1}|^2\right] & = \frac{1}{M^2}{{\frac{K_k}{K_k+1}}{\frac{1}{K_\ell+1}}\bar{\bf {h}}_k^H{\bf R}_\ell\widetilde{\bf {h}}_\ell}
\end{align}
such that ${\rm S_1}\stackrel{{\rm m.s.}}{\longrightarrow} 0$ if condition C1 is fulfilled. Clearly, the same methodology can be followed for ${\rm S_2}$. For ${\rm S_3}$, we have 
\begin{align}
	{\mathbb E}\left[{\rm S_3}\right] &= 0\\
	{\mathbb E}\left[|{\rm S_3}|^2\right] & = \frac{1}{M^2}\frac{1}{K_k+1}{\frac{1}{K_\ell+1}}{\rm tr}({\bf R}_\ell{\bf R}_k)
\end{align}
	such that ${\rm S_3}\stackrel{{\rm m.s.}}{\longrightarrow} 0$ if condition C2 is fulfilled. The proof concludes by evaluating the diagonal terms of ${\bf G}^H{{\bf G}}$, i.e., $\frac{1}{M}\left[{\bf G}^H{{\bf G}}\right]_{kk}$ which also converge in the mean-squared sense when conditions C1 and C2 are fulfilled.
\end{proof}
\end{proposition}
\vspace{-5pt}
Note that the above result holds for a very general fading model as outlined in \eqref{eq:System_model1}. Although Proposition 1 is an asymptotic result we can utilize it even for a finite number of antennas to replace 
${\bf G}^H{{\bf G}}\approx {\mathbb E}\left[{\bf G}^H{\bf G} \right] $ with very good accuracy \cite{HT_MMSE}. 
\color{black}
Most importantly, such a substitution can facilitate the performance analysis of massive MIMO with different linear precoding/detection schemes in which the random term ${\bf G}^H{{\bf G}}$ appears very often \cite{MAMI_book}.

%\end{etaremune}

%\newpage
%Referring to \eqref{eq:System_model1}, we recall the following properties \cite{Hoydis_thesis}:
%\begin{align}
%||\bar{\bf {h}}_k||^2 = \bar{\bf {h}}_k^H\bar{\bf {h}}_k = M \label{eq:prop1}\\
%\frac{1}{M}\widetilde{\bf {h}}_k^H\widetilde{\bf {h}}_\ell \stackrel{{\rm a.s.}}{\longrightarrow} 0 \label{eq:prop2}\\
%\frac{1}{M}\widetilde{\bf {h}}_k^H\widetilde{\bf {h}}_k\stackrel{{\rm a.s.}}{\longrightarrow} 1 \label{eq:prop3}\\
%\frac{1}{M}\widetilde{\bf {h}}_k^H{\bf A}\widetilde{\bf {h}}_k -\frac{1}{M}{\rm tr}({\bf A})\stackrel{{\rm a.s.}}{\longrightarrow} 0. 
%\end{align}
\vspace{-6pt}
\color{black}
\begin{figure*}[t!]
    \centering
    \begin{subfigure}[t]{0.3\textwidth}
        \includegraphics[width=\textwidth]{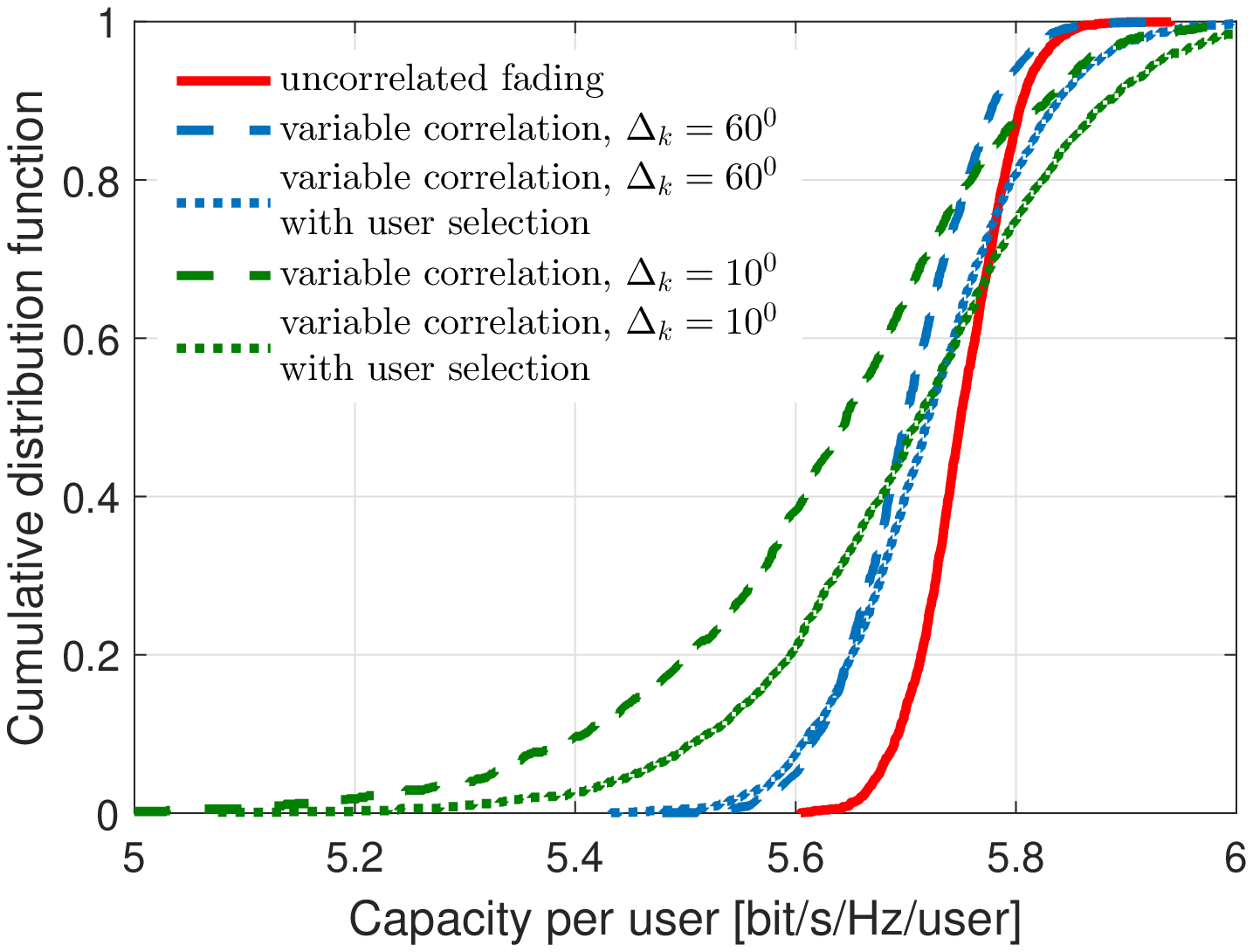}
        \caption{}
        \label{fig:1}
    \end{subfigure}
    ~ %add desired spacing between images, e. g. ~, \quad, \qquad, \hfill etc. 
    \begin{subfigure}[t]{0.3\textwidth}
        \includegraphics[width=\textwidth]{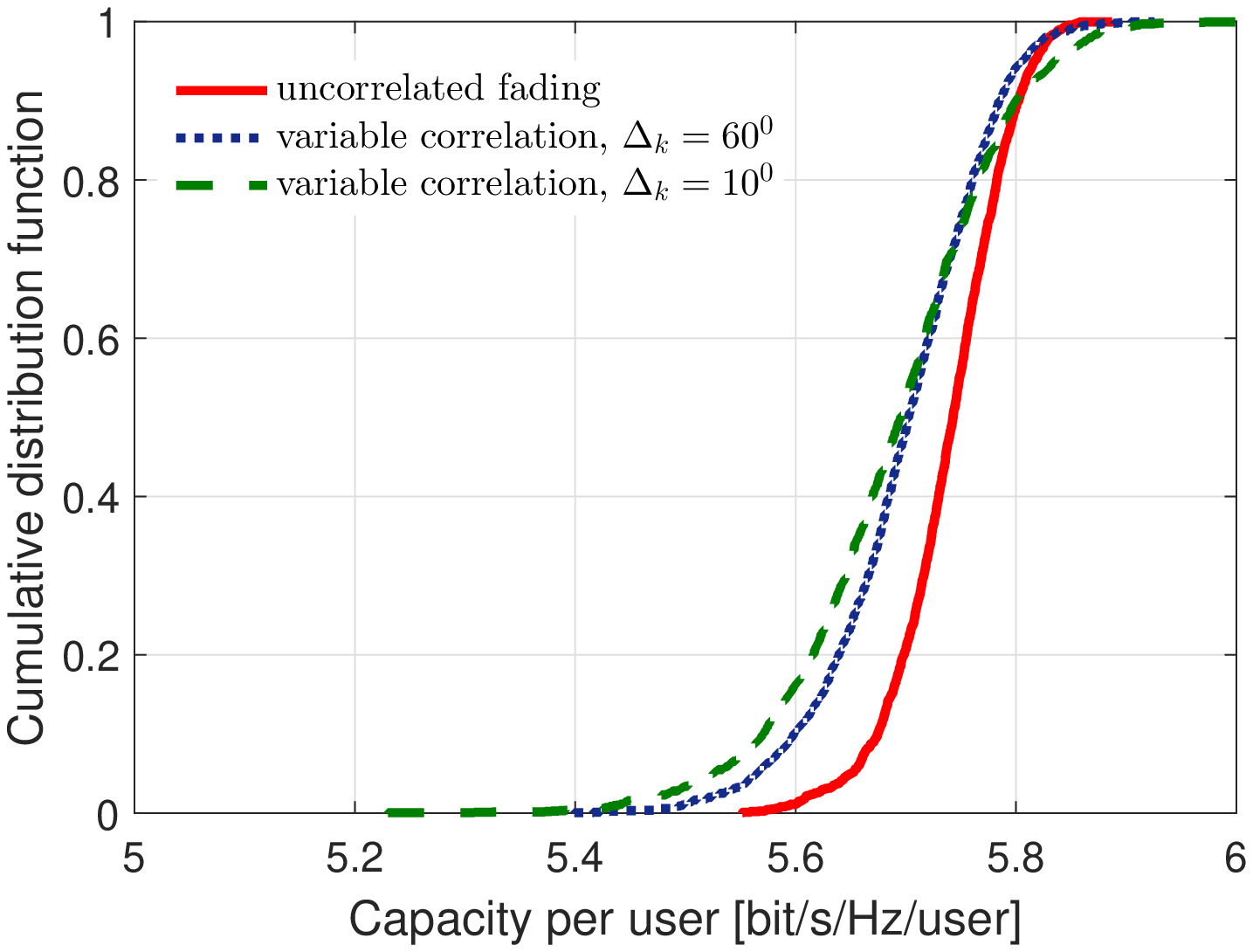}
        \caption{}
        \label{fig:2}
    \end{subfigure}
  \begin{subfigure}[t]{0.33\textwidth}
        \includegraphics[width=0.98\textwidth]{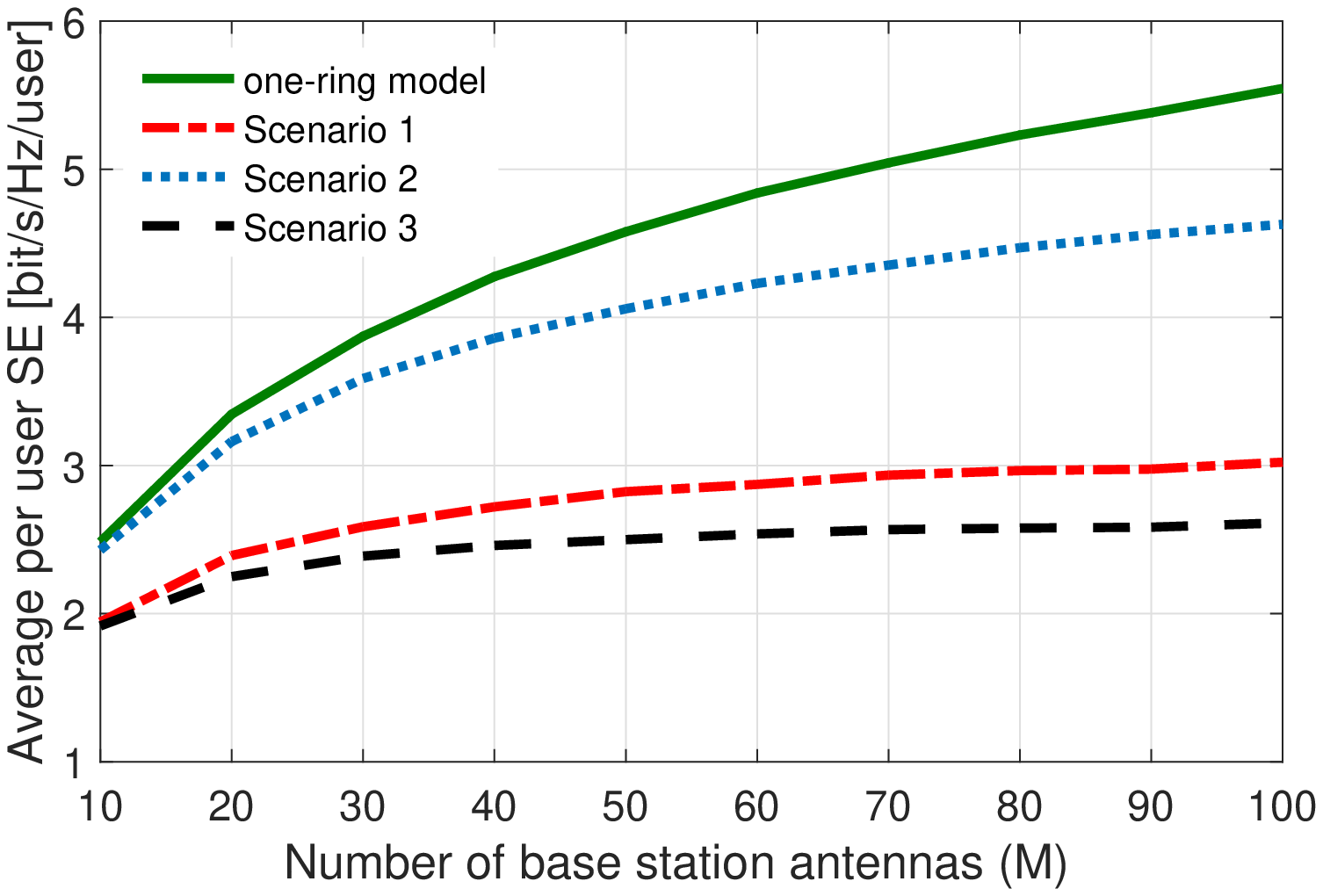}
        \caption{}
        \label{fig:3}
    \end{subfigure}
		\caption{(a): Capacity per user CDF for non-LoS channels, i.e., $K_k=0$; (b): Capacity per user CDF for random $K_k$; (c): Average SE per user with maximum ratio combining (MRC).}
\end{figure*}

\section{Numerical Results}
We now provide numerical results to verify our analysis. Our first performance metric is the capacity per user:
\color{black}\begin{align} 
C = \frac{1}{L}\log_2\det\left({\bf I}_L + p_u{\bf B}^H{{\bf B}}\right),
\end{align}
where $p_u$ is the normalized transmit power and ${\bf B}={\bf G}{\bf D}^{1/2}$, where ${\bf D}$ is the $L\times L$ diagonal matrix containing the large-scale fading coefficients, which are generated using the model of \cite{JSAC_paper}. \color{black}  To explore the effects of ${{\tt term1}}$--${{\tt term4}}$, we consider two special cases for the covariance matrix:
\begin{itemize}
\item Case 1 (uncorrelated fading): ${\bf R}_k = {\bf I}_M$, for all $k=1, \ldots, L$.

\item Case 2 (variable
correlation): We use the one-ring correlation model \cite{HT_MMSE}. With the one-ring correlation model, the $(i,j)$-th entry of ${\bf R}_k$ is given by
\begin{align}
 [{\bf R}_k]_{i,j}= \frac{1}{2\Delta_{k}}\int_{-\Delta_{k} + \phi_0^k}^{\Delta_{k}+ \phi_0^k} e^{-j\frac{2\pi d}{\lambda}(j-i) \sin(\phi_k)}d\phi_k,
\end{align}
where $\Delta_{k}$ is the azimuth angular spread corresponding to the $k$-th user, $\phi_0^k$ is the nominal direction-of-arrival, $\lambda$ is the wavelength, and $d$ is the antenna spacing. Furthermore, the LoS component is modeled as in \eqref{eq:st_vec}.
%\begin{align*}
	%\bar{\bf {h}}_k = \left[1,e^{-j\frac{2\pi d}{\lambda}\sin{(\phi_0^k)}}, \cdots,e^{-j\frac{2\pi d}{\lambda} (M-1)\sin{(\phi_0^k)}}\right]^T.
%\end{align*}

\end{itemize}

For now, we choose $M=100$, $L=10$, $p_u=0$dB, and half-wavelength antenna spacing. In addition, the $L$ angles $\{\phi_0^k\}$ 
are i.i.d. uniform random variables, distributed in $[0, 2\pi]$ while ${\bf D}={\rm diag}[0.749, 0.546, 0.425,
0.635, 0.468,  0.31, 0.64, 0.757,\\ 0.695, 0.515]$. Figure 1(a) shows the cumulative distribution  of the capacity per
user for $K_k=0$ and different azimuth angular spread (in degrees). With $K_k=0$, the channel does not have the LoS component, and hence, the effect of ${{\tt term2}}$ can be exclusively exploited. We can see that the capacity for the practical case (variable correlation)  is very close to the one for the ideal case (uncorrelated fading, i.e., ${\bf R}_{k} = {\bf I}_{M}$) especially in the high CDF tail. Yet, we can see that in the low CDF tail there is still a noticeable performance gap between these cases, especially when $\Delta_k=10^o$. This comes from the effect of ${{\tt term2}}$. The dotted curves (with user selection) represent
the cases that two users which cause the largest ${{\tt term2}}$ are dropped from service. We  see that after dropping two users from service, the performance gap between Case 1 and Case 2 reduces significantly. This suggests that we need to drop a small number of users from service to make massive MIMO working as in the ideal case. 

We next consider the case which includes both LoS and non-LoS components to examine the effects of all terms ${{\tt term1}}-{{\tt \tt term4}}$. Figure 1(b) shows the cumulative distribution of the capacity per user for random $K_k$ and different angular spreads. The values of $K_k$ are randomly and independently chosen such that they are uniformly distributed in $[0, 2]$.  Due to the presence of all four terms, the effect of the azimuth angular spread is not that significant compared to Fig. 1. 

\color{black}
Finally, Fig. 1(c) shows the average-per-user spectral efficiency for an MRC detector \cite[Eq. (6)]{Masouros} as a function of $M$. We consider a two-user network with $\Delta_k=60^o$, ${\bf D}={\rm diag}[0.749, 0.546]$, $K_1=K_2=1$, whereas all other parameters are kept the same. In line with our theoretical analysis, we study three detrimental scenarios: in Scenario 1 we assume that $\bar{\bf {h}}_2=\sqrt{M}{\bf u}_1^{(1)}$ ; in Scenario 2 we assume that ${\bf R}_1={\bf R}_2={\rm diag}[M/2,M/(2M-2),\ldots,M/(2M-2)]$ and in Scenario 3, we assume that $\bar{\bf {h}}_1=\bar{\bf {h}}_2$. All three scenarios make the SE saturate with $M$, whilst the most destructive scenario is when the LoS responses are aligned. 
\color{black}

%\begin{figure}[t]
    %\centerline{\includegraphics[width=0.48\textwidth]{fig1.eps}}
    %\caption{Capacity per user CDF for non-LoS channels, i.e., $K_k=0$. }
		%%Here $M=100$, $L=10$, and $p_u=0$ dB.}
    %\label{fig:1}
%\end{figure}
%
%\begin{figure}[t]
    %\centerline{\includegraphics[width=0.48\textwidth]{fig2.eps}}
    %\caption{Capacity per user CDF for random $K_k$.} 
		%%Here $M=100$, $L=10$, and $p_u=0$ dB.}
    %\label{fig:2}
%\end{figure}
%
\vspace{-5pt}
\section{Conclusion}
\color{black} We have identified a set of scenarios under which massive MIMO can potentially fail in Ricean fading channels. These extreme scenarios require non-vanishing alignment between LoS vectors and/or between the covariance matrices and the LoS vectors. \color{black}  In case a massive MIMO system encounters such a case, any standard scheduling scheme can compensate for the performance loss by dropping the highly-correlated users. As a final remark, we point out the small variations of the capacity around its mean value across all cases in Figures 1 and 2, which corroborates the theoretical findings of Proposition 1.

\vspace{-5pt}
\bibliographystyle{IEEEtran}

\begin{thebibliography}{}
\providecommand{\url}[1]{#1}
\csname url@samestyle\endcsname
\providecommand{\newblock}{\relax}
\providecommand{\bibinfo}[2]{#2}
\providecommand{\BIBentrySTDinterwordspacing}{\spaceskip=0pt\relax}
\providecommand{\BIBentryALTinterwordstretchfactor}{4}
\providecommand{\BIBentryALTinterwordspacing}{\spaceskip=\fontdimen2\font plus
\BIBentryALTinterwordstretchfactor\fontdimen3\font minus
  \fontdimen4\font\relax}
\providecommand{\BIBforeignlanguage}[2]{{%
\expandafter\ifx\csname l@#1\endcsname\relax
\typeout{** WARNING: IEEEtran.bst: No hyphenation pattern has been}%
\typeout{** loaded for the language `#1'. Using the pattern for}%
\typeout{** the default language instead.}%
\else
\language=\csname l@#1\endcsname
\fi
#2}}
\providecommand{\BIBdecl}{\relax}
\BIBdecl

\end{thebibliography}


\begin{thebibliography}{1}
%\begin{spacing}{1.15}
%\scriptsize{

%\bibitem{MaMI} F. Rusek, \textit{et al.}, ``Scaling up MIMO: Opportunities and challenges with very large arrays,'' \textit{IEEE Signal Process. Mag.}, vol. 30, no. 1, pp. 40--60, Jan. 2013.
%
%\bibitem{Ngo1} H. Q. Ngo, E. G. Larsson, and T. L. Marzetta, ``Energy and spectral efficiency of very large multiuser MIMO systems,'' \textit{IEEE Trans. Commun.}, vol. 61, no.4, pp.1436--1449, Apr. 2013.

\bibitem{MAMI_book} T. L. Marzetta, E. G. Larsson, H. Yang, and H. Q. Ngo, \textit{Fundamentals of Massive MIMO}. Cambridge, UK: Cambridge University Press, 2016.

%\bibitem{Zhang} Q. Zhang, S. Jin, K.-K. Wong, H. Zhu, and M. Matthaiou, ``Power scaling of uplink massive MIMO systems with arbitrary-rank channel means,'' \textit{IEEE J. Sel. Topics Signal Process.,} vol. 8, no. 5, pp. 966–-981, Oct. 2014.

\bibitem{Ngo_EUSIPCO} H. Q. Ngo, E. G. Larsson, and T. L. Marzetta, ``Aspects of favorable propagation in massive MIMO,'' 
in \textit{Proc. EUSIPCO}, Sept. 2014.

\bibitem{Masouros} C. Masouros and M. Matthaiou, ``Space-constrained massive MIMO: Hitting the wall of favorable propagation,'' \textit{IEEE Commun. Lett.}, vol. 19, no. 5, pp. 771--774, May 2015.

\bibitem{Emil_PC} E. Bj\"{o}rnson, \textit{et al.}, ``Massive MIMO has unlimited capacity,'' 
\textit{IEEE Trans. Wireless Commun.}, vol. 17, no. 1, pp. 574--590, Jan. 2018. 

\color{black}
\bibitem{Luca} L. Sanguinetti, A. Kammoun, and M. Debbah, ``Asymptotic analysis of multicell massive MIMO over Rician fading channels,''
in \textit{Proc. IEEE ICASSP}, March 2017. pp. 3539--3543.

\color{black}
\bibitem{Beualieu}  X. Wu,  N. C. Beaulieu, and D. Liu, ``On favorable propagation in massive MIMO systems and different antenna configurations,'' \textit{IEEE Access}, vol. 5, pp. 5578--5593, May 2017.

\bibitem{Caire} J. Nam, G. Caire, and J. Ha, ``On the role of transmit correlation diversity in multiuser MIMO systems,'' \textit{IEEE Trans. Inf. Theory}, vol. 63, no. 1, pp. 336--354, Jan. 2017.

\bibitem{Tataria2017} H. Tataria, \textit{et al.}, ``Impact of line-of-sight and unequal spatial correlation in uplink MU-MIMO systems,'' \textit{IEEE Wireless Commun. Lett.}, vol. 6, no. 5, pp.  634--637, Oct. 2017.

\bibitem{GLOBECOM} J. H. Chen, ``When does asymptotic orthogonality exist for very large arrays?'' in \textsl{Proc. IEEE GLOBECOM}, Nov. 2013, pp. 6--10.

\bibitem{HT_MMSE} H. Tataria, \textit{et al.}, ``Revisiting MMSE combining for massive MIMO over heterogeneous propagation channels,'' in \textit{Proc. IEEE ICC}, May 2018.

\color{black}
\bibitem{JSAC_paper} H. Q. Ngo, \textit{et al.}, ``Multipair full-duplex relaying with massive arrays and linear processing,'' \textit{IEEE J. Sel. Areas Commun.}, vol. 32, no. 9, pp. 1721--1737, Oct. 2014. 

\color{black}

%\bibitem{Hoydis_thesis} J. Hoydis, Random Matrix Methods for Advanced Communication Systems, Ph.D. dissertation, SUPELEC, April 2012.


\end{thebibliography}

%\IEEEtriggeratref{21}     

\end{document}